\documentclass[aip,
 amsmath,amssymb,
 preprint,%
]{revtex4-1}

\usepackage{graphicx}
\usepackage{dcolumn}
\usepackage{bm}

\usepackage[utf8]{inputenc}
\usepackage[T1]{fontenc}

\begin{document}


\title{Dynamic Light Scattering based microrheology of End-functionalised triblock copolymer solutions} 

\author{Ren Liu}
\affiliation{University of Cambridge, Cavendish Laboratory,Cambridge CB3 0HE, United Kingdom}

\author{Alessio Caciagli}
\affiliation{University of Cambridge, Cavendish Laboratory,Cambridge CB3 0HE, United Kingdom}

\author{Jiaming Yu}
\affiliation{University of Cambridge, Cavendish Laboratory,Cambridge CB3 0HE, United Kingdom}

\author{Xiaoying Tang}
\affiliation{University of Cambridge, Cavendish Laboratory,Cambridge CB3 0HE, United Kingdom}

\author{Rini Ghosh}
\affiliation{University of Cambridge, Cavendish Laboratory,Cambridge CB3 0HE, United Kingdom}

\author{Erika Eiser}
\email{erika.eiser@ntnu.no}
\affiliation{University of Cambridge, Cavendish Laboratory,Cambridge CB3 0HE, United Kingdom}
\affiliation{PoreLab, Department of Physics, Norwegian University of Science and Technology, N-7491 Trondheim, Norway}

\date{\today}

\begin{abstract}
`Soft' patchy surfactant micelles have become an additional building tool in self-assembling systems. The triblock copolymer, Pluronic\textsuperscript{\tiny\textregistered} F108, forms spherical micelles in aqueous solutions upon heating leading to a simple phase diagram with a micellar crystalline solid at higher temperatures and concentrations. Here we report the strong influence of end-functionalising the chain ends either with an azide or azide-DNA complex on the systems´ phase behaviour. We find that the azide(N$_3$)-functionalisation renders the chain ends weakly hydrophobic at lower temperatures, causing them to self-assemble into flower-micelles. This hydrophobicity increases with increasing temperature and poses a competing self-assembling mechanism to the solvent induces hydrophobic interactions between the middle-blocks of F108 at higher temperatures and leads to a macroscopic phase separation that is absent in the pure F108 system. However, when we attached short, hydrophilic single-stranded (ss)DNA to the azide groups via click chemistry the chain ends became `sticky´ due to DNA hybridisation below the melting temperature of the complementary ssDNA ends while reverting to hydrophilic behaviour above. We characterise their structural and rheological properties via Dynamic Light Scattering (DLS) and DLS-based passive microrheology with an improved time-frequency domain inversion step. We present the structural behaviour of dilute and semi-dilute solutions of the original F108 system and compare the results with solutions containing either the F108-azide (F108-N$_3$) or partially DNA-functionalised F108-azide chains. Our DLS and microrheology studies inform us on how the attachment of azide groups on F108 changes the mechanical and structural properties of micellar fluids pioneering further characterisation and design of these hybrid systems.
\end{abstract}



\maketitle 

\section{Introduction}
The self-assembly of block copolymers into microphase-separated structures is a research topic of high relevance in polymer science and various practical applications in nanoscience and biotechnology \cite{batrakova2008pluronic,lee2008recent,patra2018nano}. Depending on their block-size ratio, copolymers can self-assemble into classical shapes such as spheres, cylinders and bilayers \cite{zhang1995multiple}, but they are not suitable as building blocks for higher-order assemblies. Recently, patchy micelles have been prepared based on directed self-assembly of ABC type triblock copolymers forming multicompartment micelles to achieve superstructures  \cite{groschel2012precise}. This has greatly expanded the concept of `patchy' colloids, as the building subunit can now be tuned from soft to hard matter, and hybrid systems can be designed \cite{lunn2015self}, bringing a whole new dimension to the use of `patchy' colloids in the rational design of hierarchical super-structures \cite{ravaine2017synthesis, yan2016self, teixeira2017phase}.

DNA hybridization is a major player in the realization of functional and patchy particles with highly selctive interactions \cite{geerts2010, DiMichele2013}. Due to its thermo-reversibility and high specificity, it has been extensively employed to drive and direct colloidal self-assembly via temperature control \cite{feng2013dna, jones2010dna, macfarlane2011nanoparticle}. Both `hard' and `soft' colloids have been successfully functionalized with DNA \cite{mirkin1996dna, alivisatos1996organization, valignat2005reversible, nykypanchuk2008dna, hadorn2012specific, parolini2015volume, joshi2016kinetic, zhang2017sequential}. Their extension to polymeric micelles has, however, not been attempted yet. Recent advances have shown the possibility of attaching DNA to PEG-PPO-PEG triblock copolymers \cite{caciagli2018dna, lee2018dna}. These copolymers, commercially known as Synperonic (manufactured by Croda), Pluronics\textsuperscript{\tiny\textregistered} (BASF) or Poloxamers (ICI), have been successfully used in several applied areas such as particle synthesis \cite{wang2009block}, emulsion formulation \cite{tadros2015viscoelastic}, cosmetics \cite{agnely2006properties} and coatings\cite{chang2010systematic}. The temperature-responsive, non-toxic nature of Pluronics\textsuperscript{\tiny\textregistered} marks them `smart' polymers that actively respond to changes in the surrounding environment, rendering them excellent candidates for applications in drug delivery \cite{almeida2012pluronic, bodratti2018formulation}. By combining their intrinsic thermal response to the specific and thermo-reversible DNA hybridization mechanism, novel structures with exotic phase diagrams can be envisioned. In particular, it could be possible to fine-tune the rheological properties of the underlying system in a similar way to DNA-based hydrogels \cite{fernandez2018microrheology, xing2011self} but with the advantages of better scalability. This opens the possibility for a large number of applications related to biomedicine and drug delivery, in which fine control of the materials´ mechanical properties is required.

Here we present a first study on the aggregation and phase behaviour of the N$_3$ and N$_3$-DNA functionalised triblock copolymer F108, which is known to form spherical micelles upon heating. We used a simple protocol to first attach the azide groups to the free PEO-ends of F108, of which a fraction was reacted to single-stranded DNA oligonucleotides via click-chemistry. The structural and rheological properties of the aqueous solutions of these systems were then studied as function of concentration and temperature via Dynamic Light Scattering (DLS) and DLS-based passive microrheology. The methodology for the latter is based on DLS in the single scattering limit, enabling us to measure the viscoelasticity of our system over a frequency range of $1-10^{5}\,\text{s}^{-1}$. Our results show an unexpected phase behaviour that emerged from the presence of free PEO-azide ends that showed attractive, non-specific interactions between each other, which became stronger with concentration and temperature, enriching the overall Pluronics phase diagram.

\section{EXPERIMENTAL SECTION}
\subsection{Materials}
Synperonic F108 (PEG$-$poly(propylene oxide) (PPO)$-$PEG), dichloromethane (DCM), triethylamine, 4-Toluenesulfonyl chloride (TsCl), diethyl ether, NaN$_3$ and solvents (purity $> 95$\%) were purchased from Sigma-Aldrich. Here we will refer to the triblock-copolymer simply as Pluronic F108 or F108 alone. The DNA single strands (ssDNA) were obtained from Integrated DNA Technologies (IDT).

\subsection{Sample Preparation}
\subsubsection{Functionalization of F108 with azide groups (N$_3$-PEG-PPO-PEG-N$_3$)}
We covalently attached N$_3$-groups to the PEO chain ends of F108, following a protocol described by Caciagli \textit{et al}. \cite{caciagli2018dna}. This protocal was adapted from a method introduced by Oh \textit{et al}. \cite{oh2015high}. The yield of this azide-functionalization was $>$\,90\%. The final product, F108-N$_3$, was dried under vacuum and stored in a freezer.
 
\subsubsection{Dibenzylcyclooctane (DBCO)-DNA preparation}
The complementary DNA strands used were:  amine-5$'$-TTT TTT TTT TTT TTT GGT GCT GCG-3$^{\prime}$, called \textit{A} and  \textit{A'}, amine-5$'$-TTT TTT TTT TTT TTT CGC AGC ACC-3$^{\prime}$.  Both have a 15 thymine (T) long, non-binding spacer. The molecular weight of the DNA strands is $ M_{\text{w}}$ = 7512\,${\text{gmol}^{-1}}$.  Here, we followed the protocol of amine-to-dibenzylcyclooctane (DBCO) functionalization described by Zupkauskas  \textit{et al.} \cite{zupkauskas2017optically}. The DBCO-DNA was then upconcentrated using a centrifugal filter (Amicon Ultra-0.5 Centrifugal Filter, MWCO = 3000\,$\text{gmol}^{-1}$) to a maximum concentration of 5\,$\text{mmolL}^{-1}$. The DBCO-DNA was kept frozen in 10mM phosphate buffer (PB) at 0.05\,mmolL$^{-1}$ until needed.

\subsubsection{DNA functionalized F108 preparation}
100 ${\mu}l$ F108-N$_3$ was mixed with 500\,${\mu}l$ of DBCO-DNA and 400\,${\mu}l$ of 10\,mM phosphate buffer. The solution was reacted at 65\,$^{\circ}$C via a strain-promoted alkyne-azide click reaction (SPAAC), while vigorously shaking for 24 hours. No further washing or centrifugation of the resultant solution was required. Beside the challenges in the realisation (which would require dialysis or centrifugation of composites with similar molecular weight) we reasoned that the SPAAC reaction achieved very high yields ($\geq$ 90\%) and we performed the functionalisation step in small volumes. Hence, we assumed that most ssDNA was covalently linked to F108-N$_3$ ends and only a negligible amount of ssDNA remained free in solution. All F108-N$_3$ samples, presented here, were prepared with a 1:1 ratio of F108-\textit{A}-DNA and F108-\textit{A'}-DNA, unless specified otherwise. 

\subsection{F108 Phase Diagram Study}
2\,cm wide glass vials were used for the sample preparation. F108 flakes were dissolved in 10\,mM PB buffer with 100\,mM NaCl at 4\,$^{\circ}$C, for which F108 is fully soluble in water. Final concentrations of 5-25\,\%w/v F108 solutions were prepared and kept refrigerated while equilibrating for 3 days. 
Visual observations at different temperatures were carried out by immersing sealed samples in either a water or oil bath, for temperatures varying between 5 and 85\,$^{\circ}$C. Samples were equilibrated for at least 30\,min at each temperature before a visual inspection was performed to determine their respective phase state.

\subsection{Dynamic Light Scattering}
DLS measurements were performed using a Malvern Zetasizer, Nano ZS (633\,nm laser). Low-volume measurements were preformed using $40\,\mathrm{\mu L}$ microcuvettes (Malvern ZEN0040). Samples were equilibrated for 20\,min at each temperature to ensure thermal and hydrodynamic equilibrium prior to each run. The duration of each run was adjusted automatically according to the estimated relaxation time of the respective sample (i.e. the time required for $g^{(2)}(\tau)$  to decay to 1) and varied between 100\,s and 30\,min. An average over three consecutive runs constitutes a measurement.

\subsection{DLS-Based Microrheology}
We employed 230\,nm polystyrene spheres coated with polyethylene glycol (PS-PEG) as tracer particles, which were provided by Cambridge Bespoke Colloids, UK. Sample evaporation was prevented by adding a thin layer of silicone oil (50\,cSt) in the cuvette on top of the sample (measurements were unaffected by this addition).
To ensure the detection of only single-scattering events, the setup was operated in a non-invasive backscattering (NIBS) mode at a scattering angle of 173$^{\circ}$. In this setting, optimal scattering was achieved for a particle volume fraction of 0.03\%, ensuring that the probe scattering dominated over direct scattering from the sample, accounting for over 90\% of the signal. We verified that the colloids did not interact with each other by performing the DLS measurements under the same experimental conditions, but different concentrations\cite{stoev2020role}. 
 
An in-house developed constrained regularisation (CONTIN) method was used to analyse the measured scattering intensities. The raw intensity-autocorrelation functions $g^{(2)}(q, \tau)$) were normalized and then converted  into electric-field autocorrelation functions $g^{(1)}(q, \tau) = g^{(1)}(\tau) $ with Matlab routines. After fitting the $g^{(1)}(\tau)$ curves using another in-house routine, we extracted the corresponding mean-squared displacements (MSDs). From these MSDs we extracted the viscosities of our samples in the zero-shear limit. Further, by Fourier transforming the MSDs we also obtained the elastic and viscous moduli, $G'(\omega)$ and $G''(\omega)$, using the generalised Stokes-Einstein relation \cite{stoev2020role}. A detailed explanation of our in-house developed Matlab routines will be presented in a separate paper, where they will be made available.

\section{RESULTS AND DISCUSSION}
\subsection{Design and Phase Diagram of the System}
\begin{figure*}
\includegraphics[width=16cm]{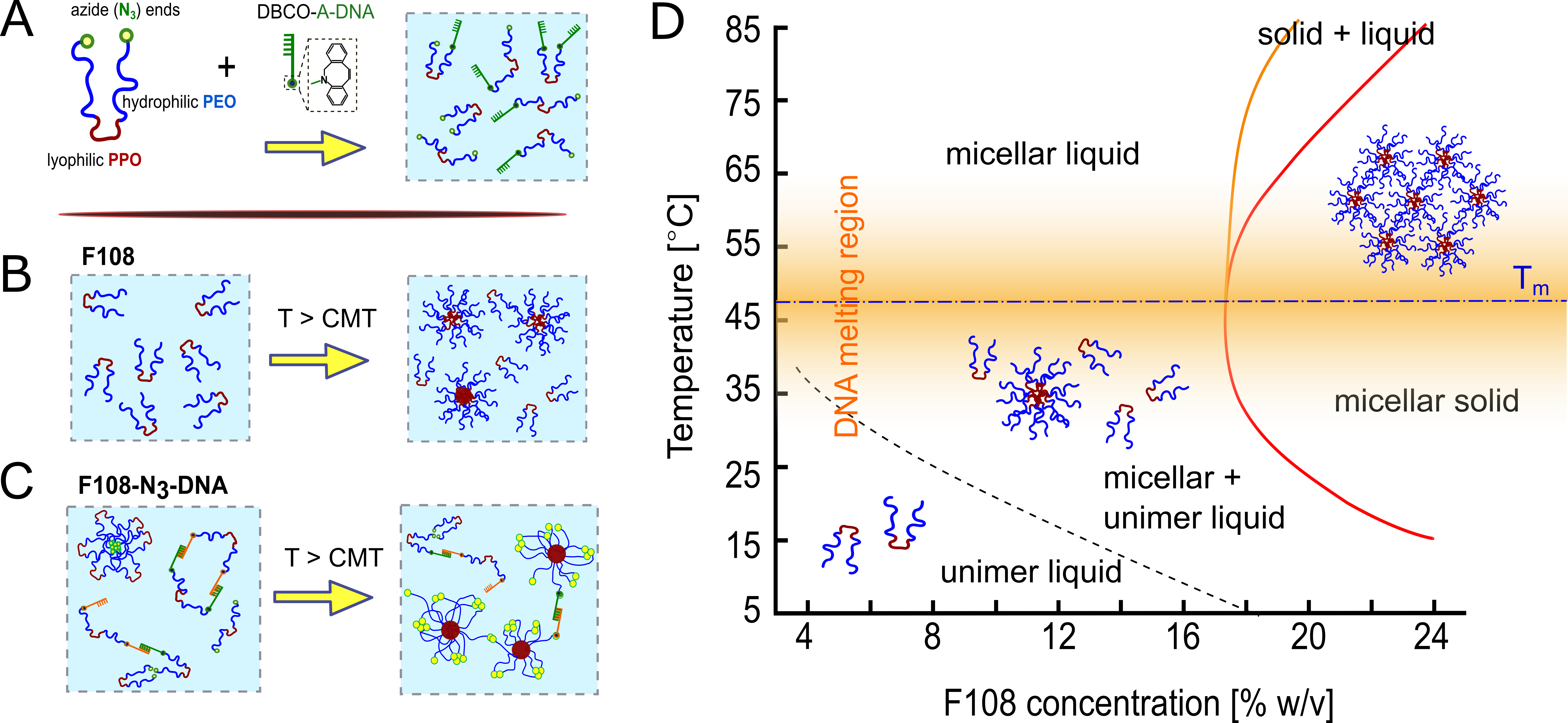}
\caption{(A) Schematic of F108 with azide-functionalised PEG-ends. N$_3$-groups were reacted to DBCO-\textit{A}-DNA with a 10:1 molar fraction, using click-chemistry. Below the CMT F108-N$_3$-\textit{A/A´} unimers are well dispersed in water. (B) Non-functional F108 chains undergo a microphase separation from unimers to spherical micelles above the CMT (dashed line)\cite{halperin1989polymeric,patel2007salt}. (C) Below the CMT chains with complementary DNA can bind forming rings or chains while the remaining F108-N$_3$-chains form transient flower micelles. Above the CMT, the unimers form standard micelles in addition to maintaining the \textit{AA´} bonds for $T < T_m \sim 48\,^{\circ}$C. (dash-dotted line); the width of the DNA melting region is indicated by the orange region in the phase diagram. (D) Experimental phase diagram of F108 in buffer solution containing 100\,mM NaCl, based on optical observations. The red line marks the transition to a single micellar solid phase while the region between the orange and red line indicates a micellar solid coexisting with a micellar liquid.}
\label{fig:exp_phase_diagram}
\end{figure*}

The phase diagram of non-functionalized F108 in aqueous solutions containing 100\,mM NaCl is presented in Figure \ref{fig:exp_phase_diagram}. F108 is a symmetric triblock copolymer of the form (EO)$_{147}$-(PO)$_{56}$-(EO)$_{147}$ with a molecular weight $M _w = 14600\,\mathrm{gmol}^{-1}$. It belongs to the hydrophilic spectrum of the Pluronics family \cite{wanka1994}. Because of their hydrophilic-to-hydrophobic size ratio F108 chains can only form spherical micelles above a critical micelle temperture (CMT) \cite{alexandridis1999solvent,salamone1996polymeric}; latter depends weakly on the F108 concentration and can be lowered substantially by adding salt \cite{diat1996,wu2006saltPluronics}. 

Aqueous F108, F108-N$_3$ and F108-N$_3$-DNA solutions are transparent. Therefore, the transition from unimer to micellar liquid can only be determined either via Small Angle X-ray Scattering (SAXS) or rheology \cite{eiser2000PRE, wu2006saltPluronics}. However, the transition from a micellar liquid to micellar solid is marked by a clear liquid to solid transition with a very narrow coexistence region. Hence, we determined the phase diagram of F108 solutions using visual inspections. Similar to earlier work by Diat \textit{et al.}\cite{diat1996}, solutions containing F108 concentrations below 18\,\%w/v remained homogeneously fluid up to 85\,$^{\circ}$C. Above $\sim$\,18\,\%w/v the samples showed a clear fluid to solid transition, which we observed by carefully tumbling the sample. Like in ref. 39, the entire samples solidified upon heating (red line in Figure \ref{fig:exp_phase_diagram}) forming a micellar crystal with face-centred cubic structure, which was verified with SAXS (these data will be presented in a separate publication). However, upon further heating, we observed partial melting (region between orange and red line in Figure \ref{fig:exp_phase_diagram}), which we attribute to chain-length polydispersity\cite{mohan2008phase}. It should be noted that Pluronics are industrial products that show large batch-to-batch variations in the CMT and liquid to solid transition boundaries. Nevertheless, the shape of the phase diagram and the micellar structure of the fluid and crystalline phases remain the same for a Pluronic material with given name.

Unlike small-surfactant systems, Pluronic suspensions show a broad unimer-to-micelle transition region \cite{erika2004tripolymer}. In Figure \ref{fig:exp_phase_diagram}D we mark this transition with a dashed line above which unimers and micelles coexist. As we will see later, our low temperature findings  for the functionalized F108 systems show a very different picture. In general, for a given concentration, the number of micelles increases with increasing temperature above the CMT until all unimers are converted into micelles. F108 micelles have an aggregation number of $\sim 50$ \cite{nolan1997light} and a hydrodynamic radius $R_H \sim 12$\,nm \cite{molino1998identification}.

ssDNA was attached to the free PEO ends by first functionalising them with a N$_3$-group with a yield above 90\,\% (Figure \ref{fig:exp_phase_diagram}A). Subsequently, DBCO-functionalised ssDNA strands were attached to the N$_3$ ends of the block copolymers via a strain-promoted alkyne-azide click reaction (SPAAC). F108-\textit{A}-DNA and F108-\textit{A'}-DNA solutions were prepared separately and then mixed in 1:1 ratios to obtain the final concentrations. In the following we refer to these mixtures as F108-N$_3$-\textit{AA'} solutions. We did not wash or centrifuge the resultant solutions because of the high yields of the click reaction. Moreover, for most experiments only a fraction of the F108-N$_3$ ends were functionalised with ssDNA; here we refer to these 10:1 molar ratios simply as F108-N$_3$-\textit{AA'} samples, unless stated otherwise. The reason for the low fraction of ssDNA per F108-N$_3$ chains was the fact that maximum yield of the click reaction could be performed only at 60\,$^{\circ}$C. Hence, ssDNA-functionalisation of all polymer-chain ends at concentrations above 18\,\%w/v would need to be performed in the micellar solid phase, which may have impeded the SPAAC reaction. Performing the reaction at lower concentrations followed by subsequent up-concentration via dialysis or centrifugation, however, is extremely costly in terms of the amount of ssDNA needed. Therefore, here we present mainly DLS studies of F108-N$_3$:F108-N$_3$-\textit{AA'} samples in the liquid state with a large fraction of unreacted F108-N$_3$ chain ends. While a simulation study on the detailed phase diagram of pure F108 and F108-\textit{AA'} solutions, in which all ends are DNA-functionalised, will be presented in a separate publication, preliminary SAXS data suggest that the azide ends are weakly hydrophobic in the unimer-region ($T < $TMC). This leads to the formation of  transient flower micelles that form and break apart rapidly. As shown in the DLS experiments the hydrophobic interactions increase both with concentration and temperature. 

Attaching DNA to the F108-N$_3$ ends introduces an additional self-assembling driving force, based on a hybridization-interaction potential that is several $k_BT$ deep for temperatures below the melting temperature $T_m$ of a given DNA duplex. $T_m$ is defined as the temperature at which half of all hydrogen-bonds between the complementary strands are formed. We measured a $T_m = 48\,^{\circ}$C at 100 mM NaCl for the \textit{AA'}-pair used here; it is well above the CMT of F108, and its melting region is indicated as orange shaded area in Figure \ref{fig:exp_phase_diagram}D. 

Thus, below the CMT the F108-N$_3$-\textit{AA'} system is expected to be a fluid suspension of rings and short connected strings of F108-N$_3$-\textit{AA'} chains coexisting with loosely connected flower micelles (Figure \ref{fig:exp_phase_diagram}C). Above the CMT the shape of the flower-micelles can no long be maintained as the increasingly bad solvent conditions for the PPO-middle-block forces them to aggregate into water free cores of regular F108 micelles. Simulataneousely, the PEO-N$_3$ chain ends tend to aggregate into small clusters. Again these chain ends can connect micelles to each other: We hypothesize that the depth of their attractive interaction potential is small at intermediate temperatures and concentrations, as the samples remain in the single phase and are completely fluid for up to about 15\% w/v. We argue that the azide-ends tend to not loop back into the water-free core of the micelles but rather stretch away as drawn in our cartoon in order to reduce unfavourable monomer-monomer interactions in the solvated PEO corona. Such a behaviour was indeed reported by Eiser \textit{et al.}\cite{eiser1999shear} who performed normal- and shear-force measurements between atomically smooth surfaces coated with telechelic chains forming a stretchered polymer-brush in good solvent conditions. These measurements also confirmed the fast dynamics between transiently connected chain ends reported in literature\cite{halperin1989polymeric}. Hence, our micelles will constantly form and break with a certain characteristic time, which depends upon the triblock copolymer species and solvent quality \cite{wang1992detection}. Well above the DNA melting temperature $T_m$, all inter-micellar DNA-bonds will be broken allowing only the azide-ends to remain in a flower-micelle type state. Of course, if all chain ends would carry complementary DNA strands we would retrieve a simple micellar liquid similar to that of the original F108 system.

\subsection{DLS and DLS-microrheology study of Semi-Dilute Solutions}
We first studied the properties of 5\,\%w/v F108 and F108-N$_3$-\textit{AA'} solutions in 100\,mM NaCl. The added salt ensured the equilibrium hybridization of the complementary DNA strands \cite{geerts2010}. We estimated that the added NaCl lowers the CMT by around 1\,$^{\circ}$C, and thus could be neglected \cite{patel2007salt, wu2006saltPluronics}. At high enough temperatures, where all F108-unimers are known to form spherical micelles we estimated their volume fraction to be $\phi \sim$ 0.03. Hence, micelle-micelle interactions can be neglected, allowing us to measure the micellar radius and eventual differences between the two systems. The raw auto-correlation curves $g^{(1)}(\tau)$ obtained by DLS showed a single-exponential decay for both systems in the temperature range 25-60\,$^{\circ}$C, as our DLS instrument did not allow us to access lower temperatures. We used a cumulant analysis method to extract the micellar hydrodynamic diameter $D_H$ and polydispersity index by fitting the short-time portion of the auto-correlation function as

\begin{equation}
\log(g^{(1)}{(\tau)}) = -\Gamma\Big(\tau-\frac{\mu_2}{2}\tau^2\Big).
\end{equation}

\begin{figure*}
\includegraphics[width=16cm]{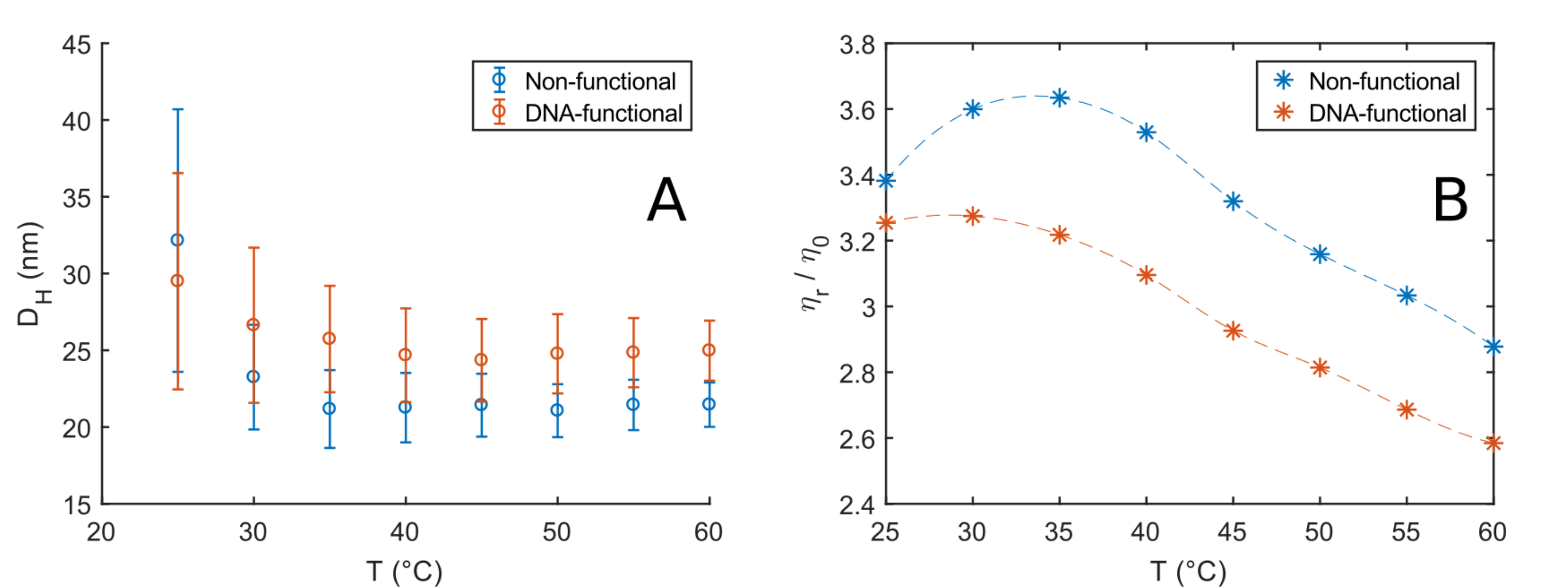}
\caption{(A) Hydrodynamic diameter $D_H$ of 5\% w/v F108 and F108-N$_3$-\textit{AA'} solutions estimated by cumulant analysis at different temperatures. The error bars represent the polydispersity index (PDI). (B) Relative viscosities, normalised by the solvent viscosity $\eta_0$ measured as function of temperature for the solutions in (A) obtained via DLS microrheology: The $\eta_r$ values were obtained from the long-time frequency region of the MSDs.}
\label{fig:Dilute_Solution}
\end{figure*}

Here, $\Gamma = q^2D$ is the decay rate, $\mu_2$ the second cummulant and $\mu_2/\Gamma^2$ is the polydispersity index (PDI). $\Gamma$ and the PDI are fitting parameters, $q$ is the scattering vector and $D$ the hydrodynamic diffusion coefficient of the scattering micelles, which was obtained via the Stokes-Einstein relation

\begin{equation}
D_H = \frac{q^2k_BT}{3\pi\eta_s\Gamma},
\end{equation}

with the solvent viscosity $\eta_s$. Latter was taken to be that of H$_2$O, as listed in the NIST database \cite{NIST}. Figure \ref{fig:Dilute_Solution}A shows the calculated hydrodynamic diameters of 5\,\%w/v solutions of F108 and F108-N$_3$-AA'; the error bars represent the PDI. Both solutions show a larger hydrodynamic size at lower temperatures, close to the CMT, which is at roughly 23\,$^{\circ}$C for F108 at this concentration. Such a behaviour is typical for Pluronics solutions \cite{zhou1988light,brown1992tribloc,nolan1997light}. It is attributed to the coexistence of unimers and micelles rapidly forming and disintegrating in the transition regime around the CMT and above. Outside the transition regime ($ T > 35\,^{\circ}$C) the hydrodynamic diameter is independent of temperature for both solutions. These findings suggest that the transition or rather coexistence region containing unimers and micelles is about 10-15\,$^{\circ}$C wide, in agreement with previous reports \cite{mortensen1992,nolan1997light}. Conversely, the scattering intensity plateaus around a maximum value for $ T > 35\,^{\circ}$C, which indicates that the number of micelles in the systems, remains constant. F108 displayed an average $D_H$ of 22\,nm, similar to earlier findings \cite{diat1996}, while the F108-N$_3$-AA' DNA solutions showed a slightly larger micellar diameter of $D_H \sim 25$\,nm. The size increase is most likely due to the presence of the ssDNA rather than the hydrophobic N$_3$-end groups that were not functionalized with DNA. The persistence length of ssDNA is of the order of 1\,nm, corresponding to about 3 bases \cite{tinland1997persistence}: Our ssDNAs are 24 bases long, and have a fully stretched length of about 8\,nm; but because they are in a very good solvent they will coil up, leading to a size that roughly matches the observed increase in the micelle diameter. Moreover, the F108-N$_3$-\textit{AA'} solutions show a slower decay to the plateau value of $D_H$ than F108, becoming temperature-independent around T $\sim 45\,^{\circ}$C. This increase in the width of the transition region is easily understood: While water is a good solvent for the PEO arms of the non-ionic F108 at low temperatures it becomes an increasingly less good solvent as $T$ increases. The addition of the negatively charged ssDNA will, however, increase the solubility of the F108 chains holding one or two DNA strands and thus shift all phase boundaries of the system to higher temperatures. Also, we do not observe any effect due to the DNA-binding below $T_m$ or binding through the hydrophobic binding between the free PEO-N$_3$-end groups that could lead to larger aggregates at these low overall concentrations.

To further elucidate these observations, we performed DLS microrheology using 200\,nm large probe particles, as explained in the experimental section. At 5\,\%w/v, both systems show little to no elasticity and can be thus characterised by their zero-shear viscosity. From the auto-correlation function

\begin{equation}
g^{(1)} (\tau) = \exp{\left(-\frac{1}{6} q^2 \langle \Delta r^2(\tau) \rangle\right)},  
\label{eq:DLSMSD}
\end{equation} 

we obtained the MSDs ($\langle \Delta r^2(\tau) \rangle$) as function of temperature. Figure \ref{fig:Dilute_Solution}B shows the viscosity curves of the two systems relative to the solvent viscosity. The trend observed for the F108 solutions is in agreement with measurements on similar systems (albeit at lower concentrations) \cite{patel2007salt,wu2006saltPluronics}. In particular, the initial upward trend in viscosity is due to the constant increase of the number of micelles in the system in the transition region. This translates to larger micellar volume fractions and a consequent increase in viscosity according to Einstein´s equation

\begin{equation}
\eta = \eta_0(1+2.5\phi_{eff}).
\label{eq:Einstein_relation}
\end{equation}

Here we considered the micelles to behave as spherical particles with an effective volume fraction $\phi_{eff}$ \cite{wu2006saltPluronics}. 

The peak at around T $\sim 35\,^{\circ}$C roughly indicates the end of the transition region for the non-functional F108 solutions, where all unimers are converted into micelles. After this point, the viscosity shows a linear decrease with increasing temperature reflecting the decreasing viscosity of the aqueous solvent. A similar but less pronounced trend was observed for the F108-N$_3$-\textit{AA'} solutions.

These viscosity results confirm our hypothesis that at these low F108-N$_3$-\textit{AA'} concentrations the DNA-attachment has tendency to increases the F108 solubility in water at all temperatures and outweighs the stickyness of the azide-end groups to connect the micelles to each other. This suggests that the total number of micelles at higher temperatures will be lower, as the fraction of F108-N$_3$ chains functionalised with DNA will most likely remain most of the time in solution. This would also explain the lower viscosity at higher temperatures for these low polymer concentrations. Although it should be noted that the average number of chains per micelle is not necessarily fixed and will be accompanied with a fast dynamics of chains entering and leaving the micelles. Hence, also the DNA-functionalised chains will occasionally enter micelles. As we will see later, the presence of the bare azide-ends becomes significant at higher concentrations and temperatures. 

\begin{figure*}
\includegraphics[width=16cm]{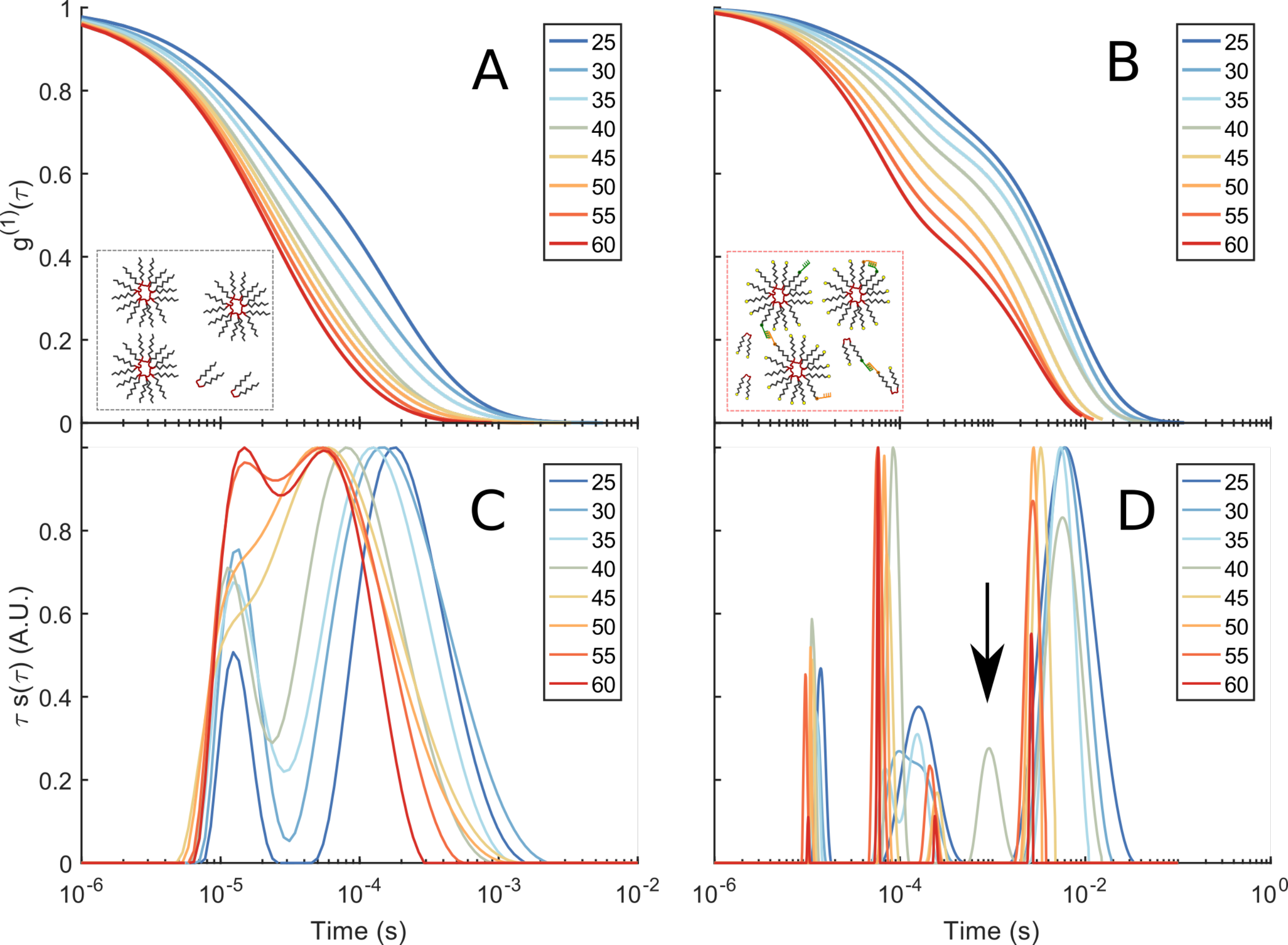}
\caption{(A) DLS spectra and corresponding relaxation times. \textit{Top}: $g^{(1)}{(\tau)}$ of a 10\% w/v F108 solution in 100\,mM NaCl measured in DLS as function of temperature.  \textit{Bottom}: Corresponding relaxation-time distributions $s{(\tau)}$, obtained by constrained regularisation fitting of the experimental $g^{(1)}{(\tau)}$ \cite{stoev2020role}. The vertical axis $\tau s(\tau)$ provides an equal area representation. The curves have been normalised to the highest peak of $s(\tau)$ to facilitate the comparison between different temperatures. (B) $g^{(1)}{(\tau)}$ of a 10\,\%w/v F108-N$_3$-\textit{AA'} solution in 100\,mM NaCl and the corresponding distributions of relaxation times. The arrow indicates the small peak at $\tau \sim 10^{-3}$\,s for $T$ = 40 \textdegree C.}
\label{fig:10perF108_compare}
\end{figure*}

\subsection{DLS and Microrheology study of Semi-Concentrated Solutions}
When doubling the concentrations of F108 and F108-N$_3$-\textit{AA'} to 10\,\%w/v in 100 mM NaCl, the CMT of the F108 system is estimated to decrease to about 21\,$^{\circ}$C \cite{alexandridis1995,patel2007salt}. At this concentration, micelle-micelle interactions can occur even though the system appeared homogeneously liquid between 5 and 80\,$^{\circ}$C. Figure \ref{fig:10perF108_compare} shows the raw autocorrelation curves $g^{(1)}{(\tau)}$ obtained by DLS for the F108 and F108-N$_3$-\textit{AA'} (1:10) solutions in a temperature range of 25-60\,$^{\circ}$C. The $g^{(1)}{(\tau)}$ curves were analysed via constrained regularisation fitting of the following form:

\begin{equation}
g^{(1)} (\tau) = g_0 +  \int_{a}^{b}\mathrm{d}\tau'\,B(\tau') e^{-\tau/\tau'}.
\label{eq:CONTINFit_Integral_multi}
\end{equation}

Plotting $\tau s(\tau)$ divided by the maximum of the individual relaxation times gives us the normalised relaxation-time curves in Figure \ref{fig:10perF108_compare}.

At 25\,$^{\circ}$C the F108 solution is in the micelle-unimer coexistence region and shows two characteristic decay times: One at $\tau_1 \approx 1.2 \times 10^{-5}$\,s that is present at all temperatures studied and a longer relaxation time $\tau_2 \approx 1.6 \times 10^{-4}$\,s with a tail up to $10^{-3}$\,s. At first, one might associate these two relaxation times with the coexisting unimers and micelles. However, using Einsteins' relation between the MSD and the diffusion coefficient of the micelles and the radius of gyration of F108 unimers ($\sim$ 5.4\,nm \cite{exerowa1997}), we estimate them to be $1 \times 10^{-5}$\,s and $0.7 \times 10^{-5}$\,s. Hence, both relaxation times fall within $\tau_1$. 

$\tau_2$ becomes a stretched exponential at higher temperatures and shifts towards faster decay times. This suggests the presence of small clusters at lower temperatures and their progressive disappearance with increasing temperature, leading to predominantly single micelles in solution. These findings are in contrast to extensive SAXS measurements by the group of Porte and others on non-functional F108 and F68, latter having the same hydrophobic to hydrophilic ratio but lower molecular weight \cite{diat1996,hamley1998,eiser2000PRE,eiserRheoA2000}. In these studies SAXS spectra taken for concentrations about 10\,\%w/v show a flat scattering spectrum in the unimer region and a typical liquid ring in the micellar region. A cluster phase could not be discerned as these would only show at even lower scattering vectors, which were not accessible at the time. Static and dynamic scattering measurements by Brown et al. \cite{brown1992} on yet other Pluronic systems showed relaxation times for unimers and micelles similar to those we find, in addition to the signature of larger, disordered aggregates at longer relaxation times. This aggregation behaviour at lower temperatures indicates the presence of weak inter-unimer and inter-micellar interactions, hinting at the fact that PEO has an upper critical solution point in addition to a lower critical solution point at around 100\,$^{\circ}$C. Indeed, when allowing the sample to settle in a glass tube for several hours and then tumbling it gently, one observes very weak Schlieren textures that disappear upon redispersion/homogenisation of the sample through vigorous stirring. 

The 10\,\%w/v F108-N$_3$-\textit{AA'} solution delivered markedly different $g^{(1)}{(\tau)}$ curves. Analysis showed that in addition to the two relaxation times observed for the pure F108 samples, a third relaxation time $\tau_3 \approx 10^{-2}$\,s appears at 25\,$^{\circ}$C. Again we see a gradual transition of $\tau_2$ to a double-exponential decay with increasing temperatures, developing all together 4 distinct decay times at 60\,$^{\circ}$C. Only at around 40\,$^{\circ}$C the system showed an additional 5\textsuperscript{th} peak at $\sim 10^{-3}$\,s. The presence of this peak is interesting, as it coincides roughly with the temperature at which all complementary DNA stands have hybridised and thus are possibly linking copolymers into small micellar chains. On increasing the temperature above $T_m = 48\,^{\circ}$C this 5\textsuperscript{th} peak disappears. To understand the presence of very long relaxation times around $\tau_3$ that persisted at all temperatures studied, we need to keep in mind that a large fraction of the Pluronic chain ends carries azide groups. We hypothesised that these groups become increasingly hydrophobic at higher temperatures and concentrations. Therefore, the azide groups will tend to form flower micelles, meaning the F108 micelles must aggregate to fulfil this assembling drive, which is expressed in the long relaxation times. But at high temperatures thermal fluctuations will also break these azide-azide interactions, hence the clusters must become smaller again, which we see in the red curves in Figure \ref{fig:10perF108_compare}D.

\begin{figure*}
\includegraphics[width=16cm]{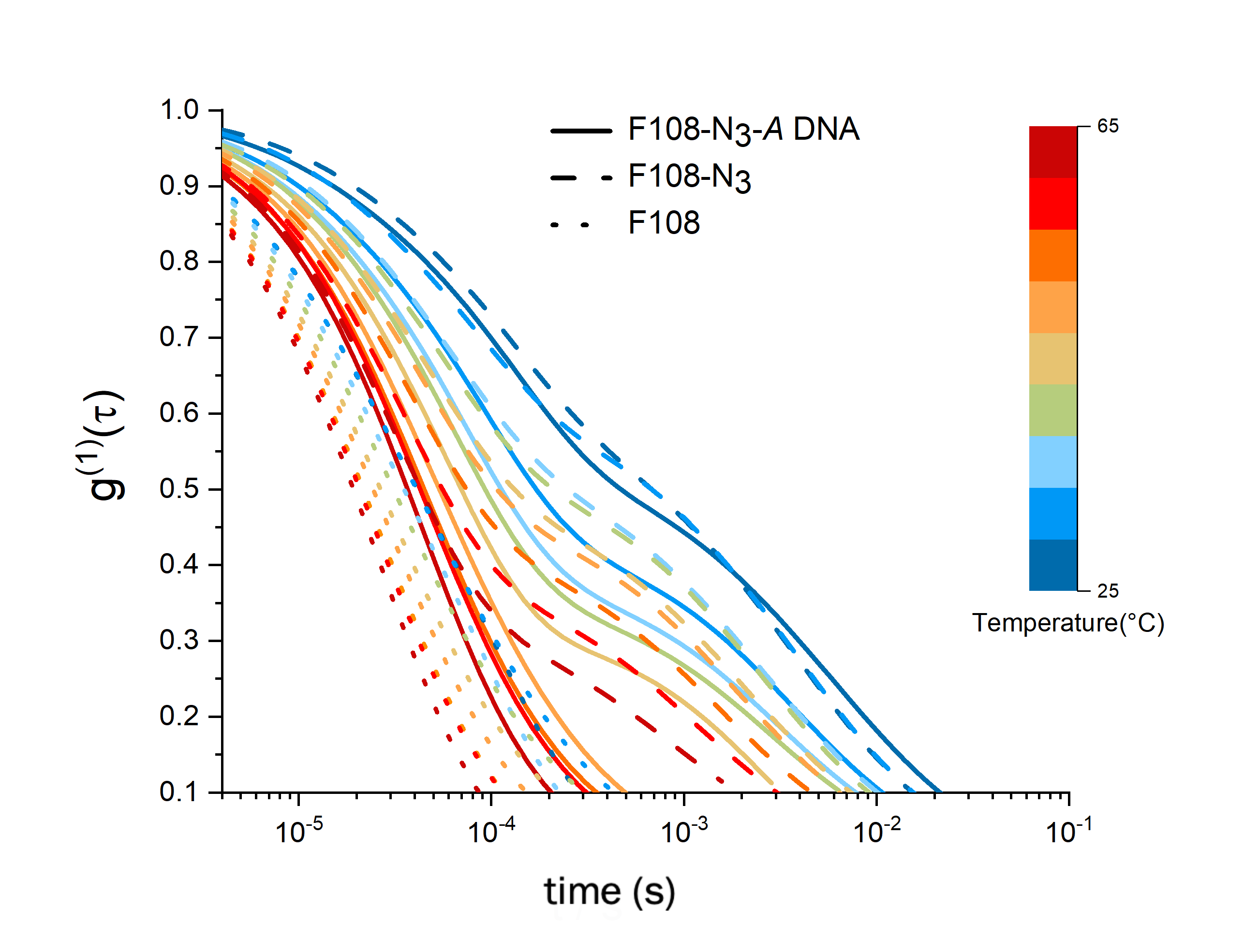}
\caption{\label{fig:8_25perF108_compare} Comparison of $g^{(1)}{(\tau)}$ spectra of a 8.25\, \%w/v F108 with a F108-N$_3$ and a F108-N$_3$-DNA (10:1) solution containing only \textit{A}-DNA such that no hybridisation can take place. All solutions were prepared in 100\,mM NaCl and measured in DLS as function of temperature.}
\end{figure*}

To this end we performed DLS studies on F108-N$_3$ samples without the DNA attachment. In Fig.~\ref{fig:8_25perF108_compare} we compare 8.25\,\%w/v samples of F108-N$_3$ with the pure F108 and F108-N$_3$-DNA (10:1) carrying only \textit{A}-DNA, not allowing for binding between PEO-chain ends. The electric field autocorrelation functions show a clear difference between the non-functionalised F108 and the one with azide and azide-DNA. In general, we observed that the $g^{(1)}{(\tau)}$ curves for F108 decay faster as the temperature increases, with no distinctly long relaxation time $\tau_3$. However, both F108-N$_3$ and F108-N$_3$-\textit{A} samples display similar $\tau_3$ relaxations at 25\,$^{\circ}$C. The similarity between the autocorrelation functions of these two azide carrying systems gradually disappeared with increasing temperature. Above 40\,$^{\circ}$C the F108-N$_3$-\textit{A} sample lost this $\tau_3$ relaxation and behaved more like the pure F108 sample. SAXS measurements of the F108-N$_3$ system indeed revealed strong aggregation with eventual macroscopic phase separation into a polymer rich solid phase and a very fluid, polymer pure region above 70\,$^{\circ}$C and 16\,\%w/v. These data will be published in a separate paper.

\subsection{Microrheology of Concentrated Solutions}
\begin{figure*}
\includegraphics[width=16cm]{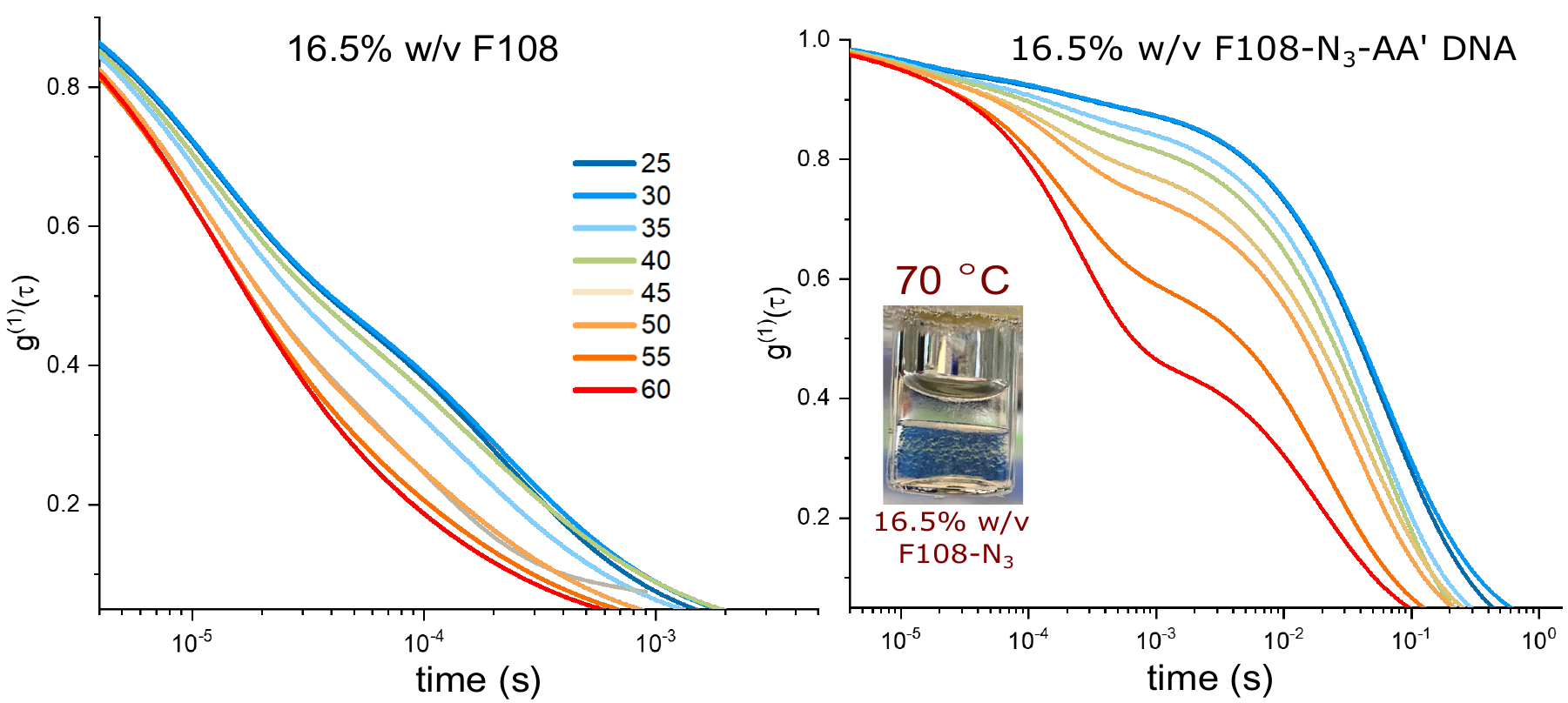}
\caption{\label{fig:16_5F108_N3} Comparison of $g^{(1)}{(\tau)}$ spectra of a 16.5\,\%w/v F108 and F108-N$_3$-\textit{AA'} (10:1) solutions. All solutions were prepared in 100\,mM NaCl and measured in DLS as function of temperature. The inset shows a photograph of a phase separated F108-N$_3$ solution heated to 70\,$^{\circ}$C}
\end{figure*}

The macroscopic phase separation in pure F108-N$_3$ solutions appeared only at $\geq$\,70\,$^{\circ}$C for concentrations larger than 16\,\%w/v and started as an overall clouding of the sample. This clouding disappeared after holding the sample at this elevated temperature for about an hour rendering the sample with two separated, clear phases (inset in Fig.~\ref{fig:16_5F108_N3}). This phase separation was thermally reversible, showing no hysteresis, as long as heating and cooling were sufficiently slow. It should be noted that the F108 samples did not phase separate but rather showed a phase transition, at which the entire F108 samples turned solid when a sufficiently high micellar concentration was reached, showing only a very narrow liquid and solid coexistence region.

Interestingly, the F108-N$_3$-\textit{AA'} solution did not show such as macroscopic phase separation, most likely again due to the increased solubility provided by the DNA even though only a fraction of the F108-N$_3$ chains had DNA attached. Nevertheless, the comparison of the autocorrelation functions for 16.5\,\%w/v, shown in Fig.~\ref{fig:16_5F108_N3}, reveals even longer relaxation times for the F108-N$_3$-\textit{AA'} solution, while those for the F108 solution do not show a drastic change, except that the overall bulk viscosity has increased, in agreement with previous measurements on F68 \cite{wu2006saltPluronics}.  

Knowing that the non-charged azide ends of the F108 chains interact with each other, we interpret the $g^{(1)}{(\tau)}$ spectra in the following way: Although we do not see a clear phase separation into a solid and liquid phase at concentrations lower than 16.5\,\%w/v of F108-N$_3$ larger clusters do form. Because of the low density mismatch between water and Pluronics, smaller clusters do remain afloat in suspension giving rise to the very long relaxation times $\tau_3$ up to about 40\,$^{\circ}$C. Above that temperature the F108-N$_3$-A DNA sample loses this long $\tau_3$ tail, showing a similar decay behaviour as the pure F108 samples. We speculate that above this temperature the aggregates forming due to the increasingly attractive azide-azide interaction leads to large enough clusters that these sediment and thus are no longer measured by the probing DLS laser beam. 

\subsection{The effect of added salt and increased DNA ratio}

\begin{figure}[h]
\centering
\includegraphics[width=13cm]{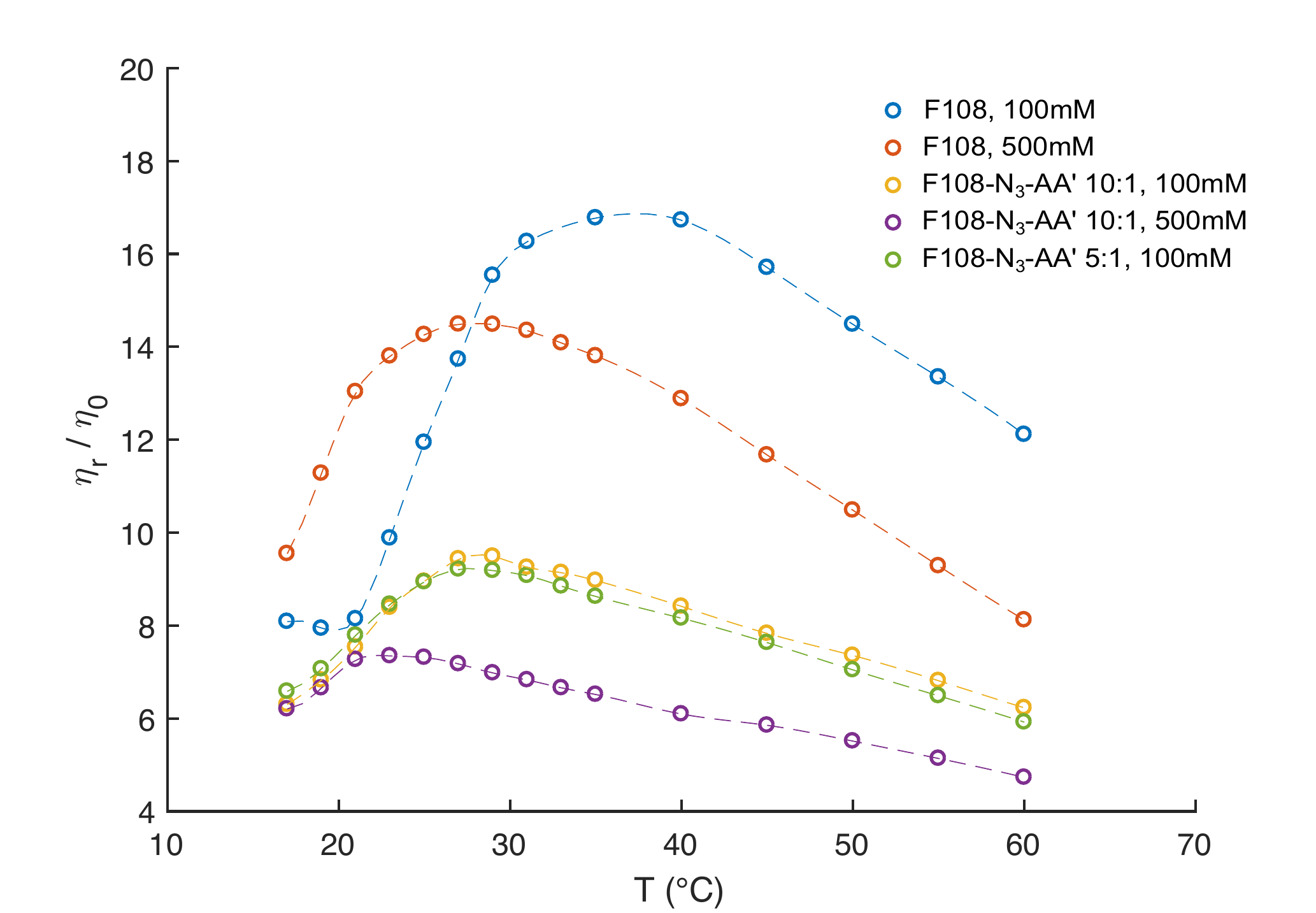}
\caption{\label{fig:relative_viscosity}  Relative viscosities of 10\,\%w/v F108, F108-N$_3$-\textit{AA'} with 10:1 and with 5:1 molar ratios as function of temperature and added NaCl, extracted from MSDs measured in DLS based microrheology.}
\end{figure}

The above observations demonstrate that the free azide ends on the F108 chains strongly influence their overall micellization and aggregation behaviour. At the same time, functionalizing the aized ends with DNA increases the solubility of the F108 chains in water, suppressing the macroscopic phase separation of the F108-N$_3$ solutions at higher concentrations and temperatures. In this context, we also examined the effect of added salt on the pure F108 and F108-N$_3$-\textit{AA'} solutions using 500\,mM NaCl, knowing that such increased salt concentrations lower the overall solubility of the PEO chains in water \cite{wu2006saltPluronics}. In addition, we investigated a F108-N$_3$-AA' DNA sample with a 5:1 molar ratio thus doubling the amount of nucleotides in the system. Performing DLS-based microrheology we show that for 10\,\%w/v solutions of the systems presented here no or very little elasticity was observed. Therefore, we could extract the zero-shear viscosities from the analysis of the MSDs, which we obtained in microrheologoy. The relative-to-solvent viscosity ratios are presented as function of temperature in Fig.~\ref{fig:relative_viscosity}.

For 100\,mM added NaCl the CMT of the 10\,\%w/v F108 solution is around $T \sim 20\,^{\circ}C$. Above that temperature we observed a strong upward trend in the viscosity \cite{wu2006saltPluronics}, following the Einstein relation linking the systems' viscosity to the increasing micellar volume fraction (see equ.(\ref{eq:Einstein_relation})). At $T \sim$ 35\,$^{\circ}C$ the viscosity reaches a maximum and then decreases with increasing temperature in a similar fashion to the semi-dilute case (Fig.~\ref{fig:Dilute_Solution}). As expected, in the presence of 500\,mM added NaCl the CMT of the F108 solution dropped below 15\,$^{\circ}$C and thus below our accessible DLS temperature range. The relative viscosity peaks around $T = 27\,^{\circ}C$ and then decreases as all possible micelles/aggregates have already formed. This behaviour is in agreement with a previous study on a similar system \cite{patel2007salt} that reports both the disappearance of large micellar clusters and the dehydration of the micelle cores. This is confirmed by the relaxation times extracted from the $g^{(1)}{(\tau)}$ curves we measured.

The DNA functionalised systems show again a remarkably different rheological response. Firstly, all viscosities are up to three times lower than those of the F108 solutions, in particular at higher temperatures. Furthermore, a clear observation of the CMT was not possible for any of the systems within our experimental setup. The reduced viscosity of these F108-N$_3$-\textit{AA'} samples is a direct result of the fact that azide and DNA binding leads to the aggregation and thus formation of a liquid of disconnected, transient clusters that is dominated by the viscosity of the aqueous background. A similar reduction in viscosity was found in solutions of DNA nanostars connected via linear DNA linkers \cite{stoev2020role}. 

\begin{figure}
\includegraphics[width=13cm]{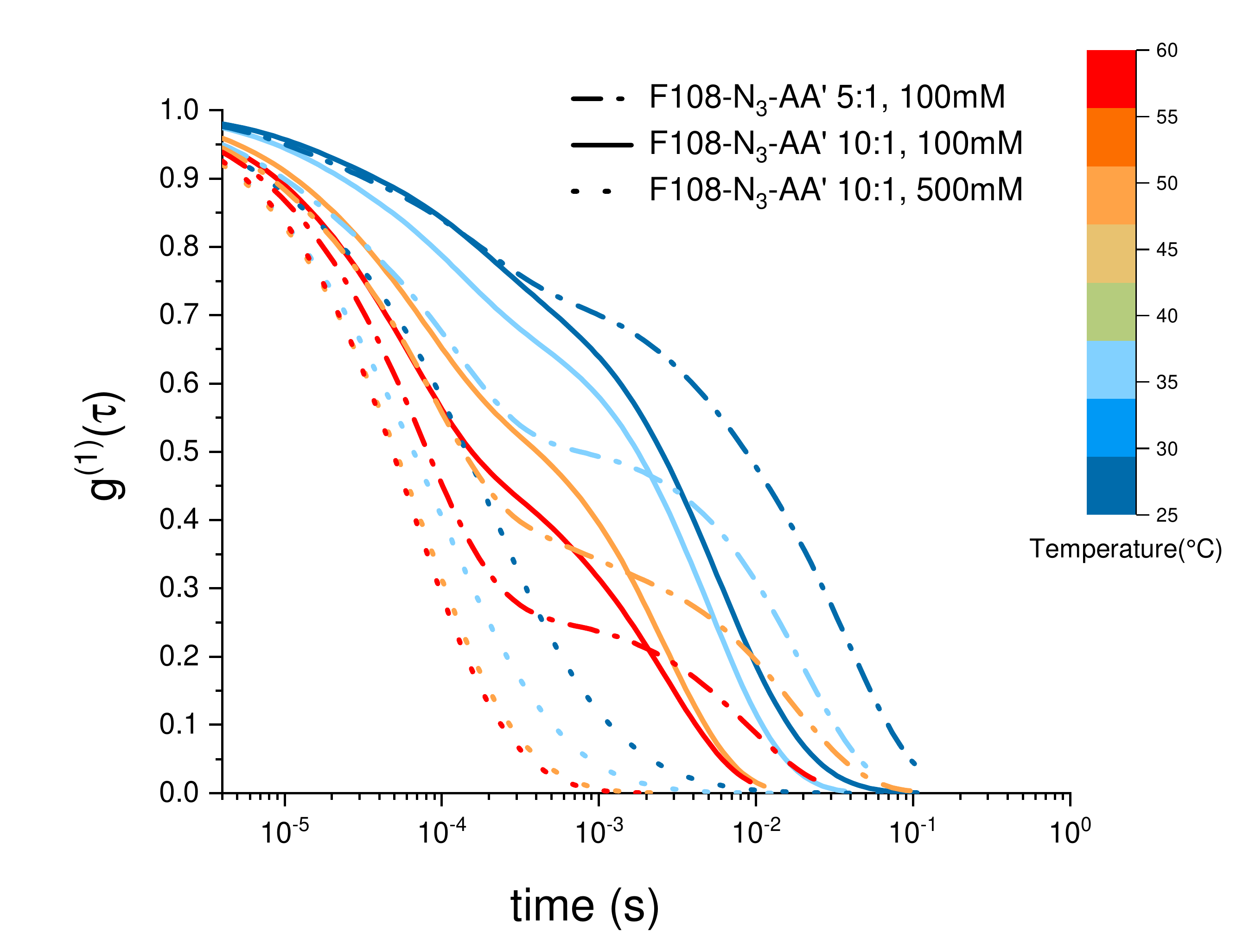}
\caption{\label{fig:merged_DLS_10perc} Electric field autocorrelation functions $g^{(1)}{(\tau)}$ of 10\,\%w/v F108-N$_3$-\textit{AA'} with 10:1 and with 1:5 molar ratios as function of temperature and added NaCl measured in DLS.}
\end{figure}

Interestingly, the two F108-N$_3$-\textit{AA'} samples measured for 100\,mM added NaCl reveal a viscosity maximum at around $T \sim 27\,^{\circ}$C followed by a linearly decreases with increasing temperature suggesting that doubling the DNA content was not sufficient to overcome the azide driven behaviour. Further, we do not observe appreciable changes near the DNA melting temperature $T_m = 48\,^{\circ}$. The change in the dominance of the different interactions becomes evident in the autocorrelation curves shown in Fig.~\ref{fig:merged_DLS_10perc}. There the $g^{(1)}{(\tau)}$ curves for the 5:1 molar ratio of the F108-N$_3$-\textit{AA'} show longer relaxation times than the 10:1 sample with lower amount of possible \textit{AA'}-DNA bonds at lower temperatures, while above the DNA melting temperature the azide-azide interaction between the free F108-N$_3$ ends dominates. Hence, if all F108-N$_3$ chain-ends would be terminated with DNA, the effect of the azides could be completely suppressed. Surprisingly, the presence of the higher added salt concentration seems to improve the solubility of the azide ends of the F108-N$_3$ chains to the point of even influencing the micelle formation of the triblock copolymer and thus keeping the overall viscosity change relatively small across the temperature range studied. Another possible interpretation could of course be that the additional salt makes the azide groups even more hydrophobic, which would entail the formation of closed rings of single F108-N$_3$ chains, thus preventing micelle formation or growth of larger aggregates. At this point we are not able to distinguish the two different mechanisms as these would require detailed SAXS measurements. 

\newpage
\section{CONCLUSIONS}
To summarise, we presented an experimental characterisation of the micellization and aggregation behaviour of azide and azide-DNA functionalized triblock copolymer, Pluronic F108, in dilute and semi-concentrated aqueous solutions and compared them to their non-functionalalized counterpart. We observed remarkable differences in the structural properties of the F108-N$_3$-\textit{AA'} samples in particular due to the attached azide-groups, which introduce non-specific, hydrophobic interactions between the solvated PEO chain-ends, which promotes the formation of flower micelles. The strength of this hydrophobic attraction between chain-ends appears to become stronger with increasing temperature and concentration, eventually leading to a macroscopic phase separation. This end-to-end attraction was competing with the driving force that induces self-assembly of the PPO middle block of the F108 chains and thus the formation of spherical micelles at high temperatures. These competing driving forces lead to cluster growth and phase separation in the concentration-temperature range in which the F108 system only forms fluid unimer, micellar or mixed phases. This cluster growth is experienced in form of strongly temperature dependent relaxation times extracted from DLS measurements.

The phase separation into a fluid top phase and a lower, solid-like gel phase at high temperatures and concentrations is suppressed when ssDNA strands are attached to the azide groups on the PEO-ends. Even if only 10\% of these ends carry the charged, strongly hydrophilic DNA strands we never observe visible aggregation below $\sim 18$\,\%w/v of F108 in solution. The structure of the various phases was studied in detailed SAXS measurements will be published separately. In order to understand the effect of the DNA on the phase behaviour of our systems, we also carried out coarse-grained simulations, where we are able to explore the entire phase diagram of the plain, aqueous F108 solutions and those in which all Pluronic chain-ends carry sticky ssDNA strands. These results will be published in a separate article as well. 

To conclude, this study demonstrates that DNA-functionalisation changes the mechanical and structural properties of micellar fluids and pioneers the further characterisation and design of these hybrid systems. By expanding the functionalisation protocol to Pluronics with a larger hydrophobic/hydrophilic ratio, micellization of the DNA-functional copolymers could be achieved as well, resulting in a binary system of non-functional and functional micelles. By controlling the interplay between these two species, a fine control over the rheological, temperature-dependent response could be obtained. The biocompatibility and thermoreversible character of these materials opens the possibility for a large number of applications related to biomedicine and drug delivery fields in which a fine control of the materials´ mechanical properties is required. 

\section{Conflicts of interest}
There are no conflicts to declare.

\section{Acknowledgments}
R.L. and J.Y. acknowledge financial support from Cambridge Trust and China Scholarship Council (CSC). A.C. and E.E. thank the ETN-COLLDENSE (H2020-MCSA-ITN-2014, Grant No. 642774) and the Winton Program for Sustainable Physics. E.E. thanks Porelab a center of excellence financed by the Research Council of Norway (RCN 262644).

\section{Data Availability}
The data that support the findings of this study are available from the corresponding author upon reasonable request.

\bibliography{F108_DNA.bib}

\end{document}